\pdfoutput=1
\RequirePackage{fixltx2e}
\documentclass[aip,jcp,reprint,floatfix]{revtex4-1}

\usepackage{amsmath,amsfonts} %
\usepackage{wasysym} %
\usepackage[acronym]{glossaries} %
\usepackage{graphicx} %
\usepackage{dcolumn} %
\usepackage{bm} %
\usepackage{ifpdf} %
\usepackage{color} %
\usepackage{nicefrac} %
\usepackage{natbib} %
\usepackage[byname]{smartref} %
\usepackage[breaklinks, %
colorlinks, %
linkcolor=blue, %
citecolor=blue, %
urlcolor=blue]{hyperref} %
\usepackage[table]{xcolor} %
\usepackage{braket}

\newcommand{\eq}[1]{Eq.~(\ref{#1})}

\newcommand{\fig}[1]{Fig.~\ref{#1}}

\newcommand{\be}{\begin{equation}}
\newcommand{\ee}{\end{equation}}
\newcommand{\bea}{\begin{eqnarray}}
\newcommand{\eea}{\end{eqnarray}}

\newcommand{\sech}{\mathrm{sech} \,}

\ifpdf %
 \fi

\begin{document}

\title{On the inclusion of the diagonal Born-Oppenheimer correction in
  surface hopping methods}

\author{Rami Gherib} %
\affiliation{Department of Physical and Environmental Sciences,
  University of Toronto Scarborough, Toronto, Ontario, M1C 1A4,
  Canada} %
\affiliation{Chemical Physics Theory Group, Department of Chemistry,
  University of Toronto, Toronto, Ontario M5S 3H6, Canada} %
\author{Liyuan Ye} %
\affiliation{Department of Physical and Environmental Sciences,
  University of Toronto Scarborough, Toronto, Ontario, M1C 1A4,
  Canada} %
\author{Ilya G. Ryabinkin} %
\affiliation{Department of Physical and Environmental Sciences,
  University of Toronto Scarborough, Toronto, Ontario, M1C 1A4,
  Canada} %
\affiliation{Chemical Physics Theory Group, Department of Chemistry,
  University of Toronto, Toronto, Ontario M5S 3H6, Canada} %
\author{Artur F. Izmaylov}
\affiliation{Department of Physical and Environmental Sciences,
  University of Toronto Scarborough, Toronto, Ontario, M1C 1A4,
  Canada} %
\affiliation{Chemical Physics Theory Group, Department of Chemistry,
  University of Toronto, Toronto, Ontario M5S 3H6, Canada} %

\date{\today}

\begin{abstract}

  The diagonal Born-Oppenheimer correction (DBOC) stems from the
  diagonal second derivative coupling term in the adiabatic
  representation, and it can have an arbitrary large magnitude when a
  gap between neighbouring Born-Oppenheimer (BO) potential energy
  surfaces (PESs) is closing. Nevertheless, DBOC is typically
  neglected in mixed quantum-classical methods of simulating
  nonadiabatic dynamics (e.g., fewest-switch surface hopping (FSSH)
  method). A straightforward addition of DBOC to BO PESs in the FSSH
  method, FSSH+D, has been shown to lead to numerically much inferior
  results for models containing conical intersections. More
  sophisticated variation of the DBOC inclusion, phase-space
  surface-hopping (PSSH) was more successful than FSSH+D but on model
  problems without conical intersections. This work comprehensively
  assesses the role of DBOC in nonadiabatic dynamics of two electronic
  state problems and the performance of FSSH, FSSH+D, and PSSH methods
  in variety of one- and two-dimensional models. Our results show that
  the inclusion of DBOC can enhance the accuracy of surface hopping
  simulations when two conditions are simultaneously satisfied: 1)
  nuclei have kinetic energy lower than DBOC and 2) PESs are not
  strongly nonadiabatically coupled. The inclusion of DBOC is
  detrimental in situations where its energy scale becomes very high
  or even diverges, because in these regions PESs are also very
  strongly coupled. In this case, the true quantum formalism heavily
  relies on an interplay between diagonal and off-diagonal
  nonadiabatic couplings while surface hopping approaches treat
  diagonal terms as PESs and off-diagonal ones stochastically.
\end{abstract}


\maketitle

\glsresetall

\section{Introduction}

The commonly used adiabatic representation defines nuclear dynamics on
multiple electronic surfaces that are coupled through terms resulted
from the nuclear kinetic energy operator acting on the
Born-Oppenheimer (BO) electronic
wavefunctions.\cite{Cederbaum:2004/CI} Kinetic energy coupling between
different BO electronic states gives rise to two effects disappearing
in the BO approximation: 1) Interstate (off-diagonal) derivative
couplings are responsible for transferring nuclear wavepackets between
electronic surfaces. 2) Second-order diagonal derivative terms, hereon
referred as diagonal Born-Oppenheimer corrections (DBOCs), modify the
BO PESs.\cite{Lengsfield:1986/jcp/84} Mathematically, DBOC is a
potential-like term and thus its addition to BO PES seems very
reasonable in consideration of quantum nuclear dynamics. Without DBOC,
BO approximation estimates for the system total energies are not
variational.\cite{Brattsev:1965/SSSR/570,Epstein:1966/jcp/836} In
regions of close proximity of BO PESs, DBOCs can become arbitrarily
large, and a nuclear wavepacket travelling on modified PESs (such
surfaces are usually called adiabatic surfaces) can undergo very
different dynamics compared to that on BO
PESs.\cite{Ryabinkin:2014/jcp/140} Generally, in nonadiabatic regions
for adequate modelling of true quantum nuclear dynamics in the
adiabatic representation, all terms related to potential and kinetic
energies as well as geometric phase appearing in conical intersections
must be taken into account.\cite{Ryabinkin:2014/jcp/140}

Often to address nonadiabatic dynamics in large systems mixed
quantum-classical (MQC) methods such as FSSH and Ehrenfest are
adequate and computationally
feasible.\cite{Tully:1990/jcp/1061,Tully:1998/fdd/110} Nuclear
dynamics in these methods is simplified to the classical level and is
governed by forces obtained from variously defined electronic
surfaces. A natural question in this context is whether adding DBOC to
electronic surfaces can improve the performance of these methods? A
nontrivial character of this question is related to the fact that in
MQC methods we do not have quantum nuclear wavepackets. Thus, although
DBOC is necessary for the correct dynamics of nuclear wavepackets, it
may not necessarily improve dynamics of classical particles. Indeed,
for conical intersection problems, a straightforward addition of DBOC
to BO PESs in the FSSH method was found to be detrimental for
dynamics.\cite{Gherib:2015/jctc/11} In further discussion we will
refer to this modification of FSSH as FSSH+D. Moreover, in the
Ehrenfest method, DBOC inclusion breaks down the invariance of the
approach with respect to the adiabatic-to-diabatic electronic basis
transformation. One of the reasons why DBOC is detrimental in CI
problems is the absence of explicit account of the geometric phase in
MQC methods.\cite{Gherib:2015/jctc/11} However, not all problems have
CIs that affect nonadiabatic dynamics, therefore including DBOC in
surface hopping approaches for non-CI problems may have its benefits
and has been advocated in
Ref.~\onlinecite{Akimov:2013/jctc/4959,Shenvi:2009/jcp/124117}.

Recently, Shenvi proposed an alternative to FSSH, the phase-space
surface-hopping (PSSH) method.\cite{Shenvi:2009/jcp/124117} The key
idea behind PSSH is the use of phase-space surfaces that incorporate
both DBOC and first derivative couplings. It was deemed that such
surfaces would be coupled weaker than corresponding BO PESs in FSSH.
On a few one-dimensional model systems it was shown that PSSH performs
very well and generally better than FSSH+D. Surprisingly, no
comparison of PSSH results with those of the original FSSH method has
been done. Also, the PSSH method have not been tried in situations
when DBOC is very large or diverging (e.g., conical intersections).

In this work we would like to assess whether including DBOC can
improve results of nonadiabatic dynamics in FSSH+D and PSSH methods
based on results in few representative one- and two-dimensional
models. If there are such cases which method among these two should be
preferred. The rest of the paper is organized as follows. Section II A
reviews the fully quantum formalism that gives rise to DBOC. Section
II B illustrates how FSSH, FSSH+D, and PSSH classical equations of
motion (EOM) can be rationalized within a general framework. In
Section III, nonadiabatic numerical simulations of various 1D and 2D
models are presented and show the strengths and limitations of FSSH,
FSSH+D and PSSH. Section IV concludes the work by summarizing main
results and discussing potential future challenges. Atomic units will
be used throughout this work.

\section{Theory}

\subsection{Diagonal Born-Oppenheimer correction}

To see how DBOC emerges in the exact quantum-mechanical formalism let
us start with the exact quantum mechanical molecular Hamiltonian \bea
\label{eq:molecularHamiltonian}
\hat{H}_{m}=\hat{T}_{n}+\hat{H}_{e}, \eea where $\hat{T}_{n}$ is the
kinetic nuclear energy operator and $\hat{H}_{e}$ is the electronic
Hamiltonian, the sum of the total molecular potential energy and the
kinetic electronic energy. The adiabatic representation involves the
basis of electronic functions $\{\ket{\phi_j(\textbf{R})}\}$ that
solve the electronic time-independent Schr{\"o}dinger equation (TISE)
for a fixed nuclear configuration $\textbf{R}$ \bea
\hat{H}_{e}\ket{\phi_i(\textbf{R})}=E_{i}(\textbf{R})\ket{\phi_i(\textbf{R})}.
\eea Using $\{\ket{\phi_j(\textbf{R})}\}$, an eigenfunction of
$\hat{H}_{m}$ can be written as \bea
\label{eq:BHexpansion}
\Psi(\textbf{r},\textbf{R})=\sum_{j}\phi_{j}(\textbf{r};\textbf{R})\chi_j(\textbf{R}),
\eea where nuclear counterparts $\chi_j(\textbf{R})$ can be obtained
from projecting the full TISE,
$\hat{H}_{m}\Psi(\textbf{r},\textbf{R})=\mathcal{E}\Psi(\textbf{r},\textbf{R})$,
onto the electronic basis \bea \label{eq1}
\sum_j\left[\braket{\phi_i(\textbf{R})|\hat{T}_{n}|\phi_j(\textbf{R})}+\delta_{ij}E_j\right]\chi_j(\textbf{R})=\mathcal{E}\chi_i(\textbf{R}).
\eea For the sake of simplicity we will only consider one nuclear
degree of freedom (DOF) with a nuclear mass $M$ \bea
\hat{T}_n=-\frac{1}{2M}{\nabla_{\textbf{R}}}^2, \eea the following
consideration can be straightforwardly extended to more nuclear DOF.
Due to the parametric dependency of adiabatic states on $\textbf{R}$,
the evaluation of
$\braket{\phi_i(\textbf{R})|\hat{T}_{n}|\phi_j(\textbf{R})}$ in
\eq{eq1} requires use of the chain rule in the action of the Laplacian
on $\ket{\phi_j(\textbf{R})}$ \bea
&&\braket{\phi_i(\textbf{R})|\hat{T}_{n}|\phi_j(\textbf{R})} =
-\frac{1}{2M}[\delta_{ij}{\nabla_{\textbf{R}}}^2 \\ \nonumber &&+
2\braket{\phi_{i}(\textbf{R})|\nabla_{\textbf{R}}\phi_{j}(\textbf{R})}\nabla_{\textbf{R}}
+\braket{\phi_{i}(\textbf{R})|{\nabla_{\textbf{R}}}^2\phi_{j}(\textbf{R})}].\label{eq:kineticenergy1}
\eea By introducing a resolution of the identity
$\sum_{k}\ket{\phi_{K}(\textbf{R})}\bra{\phi_{K}(\textbf{R})}$ inside
the last component in \eq{eq:kineticenergy1}, the matrix elements of
the nuclear kinetic energy can be expressed as \bea
&&\braket{\phi_i(\textbf{R})|\hat{T}_{n}|\phi_j(\textbf{R})} 
=-\frac{1}{2M}[\delta_{ij}{\nabla_{\textbf{R}}}^2  \nonumber \\
&&+\braket{\phi_{i}(\textbf{R})|\nabla_{\textbf{R}}\phi_{j}(\textbf{R})}\nabla_{\textbf{R}}+\nabla_{\textbf{R}}\braket{\phi_{i}(\textbf{R})|\nabla_{\textbf{R}}\phi_{j}(\textbf{R})}  \nonumber \\
&&+\sum_{k}\braket{\phi_i(\textbf{R})|\nabla_{\mathbf{R}}\phi_k(\textbf{R})}\braket{\phi_k(\textbf{R})|\nabla_{\mathbf{R}}\phi_j(\textbf{R})}].\label{eq:kineticenergy2}
\eea For a system with two electronic states, kinetic energy matrix
operator, $\mathbf{T_{n}}$, takes the following form \bea
\label{GPLVC2D}
\mathbf{T_{n}}= -\frac{1}{2M} \begin{pmatrix}
  {\nabla_{\textbf{R}}}^2-{\mathbf{d}_{12}}^2 & \nabla_{\mathbf{R}} \cdot \mathbf{d}_{12} + \mathbf{d}_{12}\cdot \nabla_{\mathbf{R}} \\
  \nabla_{\mathbf{R}} \cdot \mathbf{d}_{21} + \mathbf{d}_{21}\cdot
  \nabla_{\mathbf{R}} & {\nabla_{\textbf{R}}}^2-{\mathbf{d}_{21}}^2
\end{pmatrix}. \eea The components
$\mathbf{d}_{12}=\braket{\phi_1(\textbf{R})|\nabla_{\mathbf{R}}\phi_2(\textbf{R})}$
and ${\mathbf{d}_{12}}^2/(2M)$ are the nonadiabatic coupling vector
(NAC) and DBOC, respectively. DBOC is a function of $\textbf{R}$ and a
diagonal element of the total molecular Hamiltonian projected in the
electronic adiabatic basis \bea\label{eq:molHmodPES} \textbf{H}_m =
\mathbf{T_{n}} +
\begin{pmatrix}
  E_1(\textbf{R}) & 0 \\
  0 & E_2(\textbf{R})
\end{pmatrix}. \eea Thus, DBOC can be summed to the BO PESs and
regarded as a second-order correction in $\hbar$ \bea
\label{eq:adiabaticPES}
\tilde{E_{j}}(\textbf{R})=E_{j}(\textbf{R})+\frac{{\mathbf{d}_{12}}^2}{2M}.
\eea We will refer to $\tilde{E_{j}}(\textbf{R})$ surfaces as {\it
  adiabatic} PESs in contrast with $E_{j}(\textbf{R})$ which are
referred to as BO PESs. Adiabatic PESs, $\tilde{E_{j}}$, have always a
larger value than corresponding BO PESs, $E_{j}(\textbf{R})$. The
difference between two types of PESs grows with the length of NAC
which is inversely proportional to the difference between electronic
energies \bea
\mathbf{d}_{12}=\frac{\braket{\phi_1(\textbf{R})|\nabla_{\textbf{R}}H|\phi_2(\textbf{R})}}{E_{2}-E_{1}},
\eea and hence, both NAC and DBOC become large in the region of close
proximity of two BO PESs.

\subsection{Surface hopping methods}

Here, we provide a uniform framework rationalizing various versions of
classical nuclear EOM used in surface hopping methods.
\paragraph{FSSH:} Let us transform the molecular Hamiltonian,
$\hat{H}_{m}$ in \eq{eq:molecularHamiltonian} to its classical
analogue $H_{m}^{\textrm{cl}}$, by converting $\hat{T}_n$ to
${{\textbf{P}}^2}/{(2M)}$, where $\textbf{P}$ is the classical nuclear
momentum. This step amounts to substituting the quantum operator
$-i\nabla_{\textbf{R}}$ by the $\textbf{P}$ variable. The resulting
molecular Hamiltonian is \bea
\label{eq:molHamiltSH1}
\hat{H}_{m}^{\textrm{cl}}=\frac{{\textbf{P}}^2}{2M}+\hat{H}_{e}. \eea
By projecting $\hat{H}_{m}^{\textrm{cl}}$ onto the adiabatic basis we
obtain \bea
\label{eq:molHamiltSH2}
\hat{H}_{m}^{\textrm{cl}}=\delta_{ij}\left(\frac{{\textbf{P}}^2}{2M}+E_{i}(\textbf{R})\right),
\eea which corresponds to uncoupled EOM whereby nuclei are classical
and evolve on BO PESs $E_{i}(\textbf{R})$.

It should be noted that by following this route, DBOC does not emerge
due to the quantum-classical transformation that removes quantum
kinetic energy operator before the adiabatic electronic basis is
introduced.

\paragraph{FSSH+D:} An alternative route to the classical nuclear EOM
involves inverting the order of the quantum-classical transformation
and the projection to the adiabatic electronic basis. This inversion
amounts to starting from \eq{eq:molHmodPES} instead of
\eq{eq:molecularHamiltonian} and leads to the Hamiltonian
\begin{align}
  \label{eq:molHmodPES2}
  H_{n}^{1)\rightarrow2)} =\begin{pmatrix}
    \frac{\textbf{P}^2}{2M}+\tilde{E}_1(\textbf{R}) & -\frac{i\mathbf{d}_{12}\cdot{\textbf{P}}}{M} \\
    \frac{i\mathbf{d}_{12}\cdot{\textbf{P}}}{M} &
    \frac{\textbf{P}^2}{2M}+\tilde{E}_2(\textbf{R})
  \end{pmatrix}.
\end{align}
From thereon, we can remove the off-diagonal terms and obtain
uncoupled classical Hamiltonians corresponding to two electronic
states. In this alternative route, the potential on which the nuclei
are evolving are DBOC-modified potentials $\tilde{E}_{i}(\textbf{R})$.

\paragraph{PSSH:} If one does not discard the off-diagonal couplings
in the Hamiltonian $H_{n}^{1)\rightarrow2)}$ but rather diagonalizes
$H_{n}^{1)\rightarrow2)}$ to obtain an electronic basis parametrically
dependent on $\textbf{R}$ and $\textbf{P}$ \bea
H_{n}^{1)\rightarrow2)}\ket{n_{i}^{\rm
    PS}(\textbf{R},\textbf{P})}=E_{i}^{\rm PS}\ket{n_{i}^{\rm
    PS}(\textbf{R},\textbf{P})}, \eea $\{\ket{n_{i}^{\rm
    PS}(\textbf{R},\textbf{P})}\}$ is referred as \textit{phase-space
  adiabatic} representation. $E_{i}^{\rm PS}$ are phase-space total
energies and for a two-level system they are
\begin{align}
  \label{eq:PSenergy}
  E_{\pm}^{\rm PS}(\textbf{R},\textbf{P})=& \frac{\textbf{P}^2}{2M}+\frac{1}{2}\left(\tilde{E}_1(\textbf{R})+\tilde{E}_2(\textbf{R})\right) \nonumber \\
  &\pm\frac{1}{2}\sqrt{\left({E}_1(\textbf{R})-{E}_2(\textbf{R})\right)^2+4\left(\frac{\mathbf{d}_{12}\cdot{\textbf{P}}}{M}\right)^2}.
\end{align}
For the excited state, DBOC is enhanced by the NAC related term while
for the ground state DBOC can be compensated by the NAC term. Hence,
the phase-space representation can lift the degeneracy in the
adiabatic representation. Far from strongly coupled regions,
phase-space and adiabatic representation electronic wavefunctions and
PESs converge to each other. The classical trajectories for nuclei on
a single phase-space surface $E_{\pm}^{\rm PS}$ can be obtained using
Hamilton's EOM \bea \dot{\textbf{R}}_{\pm}=\frac{\partial E_{\pm}^{\rm
    PS}}{\partial \textbf{P}}, \quad
\dot{\textbf{P}}_{\pm}=-\frac{\partial E_{\pm}^{\rm PS}}{\partial
  \textbf{R}}. \eea

In all these approaches nuclear dynamics experience stochastic hops
between PESs, the hopping probabilities are proportional to NACs and
their explicit expressions and further details on the electronic
dynamics can be found in the supplementary material.\cite{misc:supp}

\section{Numerical simulations}

To determine whether DBOC could be beneficial in surface hopping
approaches and to assess more extensively the accuracy of PSSH and
FSSH+D, we consider in this section three types of systems: 1) two
flat one-dimensional BO PES coupled nonadiabatically, 2)
one-dimensional avoided crossing model with different diabatic
couplings, 3) two-dimensional linear vibronic coupling (2D-LVC) models
containing CIs in the adiabatic representation. All FSSH and FSSH+D
simulations were performed in the adiabatic representation.

\subsection{Flat BO PESs}

\begin{figure}[!h]
  \centering
  \includegraphics[width=0.5\textwidth]{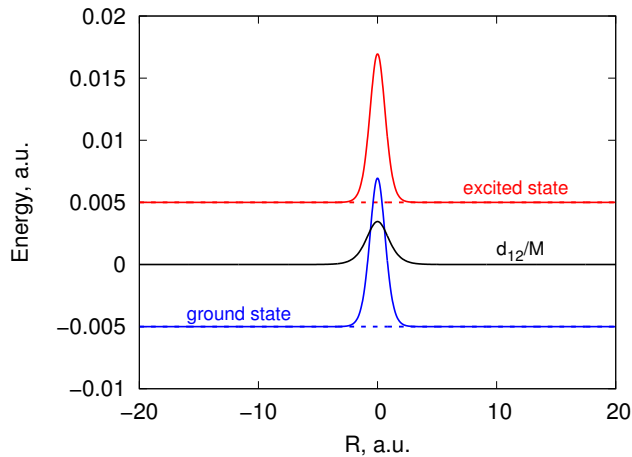}
  \caption{Nonadiabatic coupling (solid black) and PESs for the model
    with flat BO PESs: ground (dashed blue) and excited (dashed red)
    BO PESs, ground (solid blue) and excited (solid red) adiabatic
    PESs.}
  \label{fig:model_2_pes}
\end{figure}

We begin by considering model 2 of the PSSH original
paper\cite{Shenvi:2009/jcp/124117} (\fig{fig:model_2_pes}). The
molecular Hamiltonian in the diabatic representation for this model is
\begin{align}
  &H^{\rm D}_{\rm F}=-\frac{1}{2M}\frac{\partial^{2}}{\partial
    {R}^{2}}\cdot\mathbf{1}_{2}+\begin{bmatrix}
    -A\cos(\theta) & A\sin(\theta) \\
    A\sin(\theta) & A\cos(\theta)
  \end{bmatrix},
\end{align}
where $\theta=C\pi\left(\tanh(D{R})+1)\right)$, $A=0.005$, $C=5.5$,
$D=0.8$ and $M=2000$ a.u. The model involves two flat BO PES coupled
with NAC that has a $\sech^{2}({R})$ form.

The exact quantum dynamics simulations were performed using the split
operator method on a grid of 2048 points inside a box of length 40
a.u. and a time-step of 0.1 fs. The initial wavepacket was a Gaussian
$\Psi({R},0)=e^{i\braket{{P}}{R}}e^{-\left(({R}-\braket{R})/\sigma\right)^2}$
with a width parameter $\sigma=20/\braket{{P}}$. The SH simulations
were done with 2000 trajectories for all three SH methods, and
time-steps of 0.025 fs, 0.025 fs, and 0.01 fs for FSSH, PSSH, and
FSSH+D, respectivley. The initial distribution of positions and
momenta for the SH simulations was taken as the Wigner transform of
$\Psi({R},0)$.

\begin{figure}[!h]
  \centering
  \includegraphics[width=0.5\textwidth]{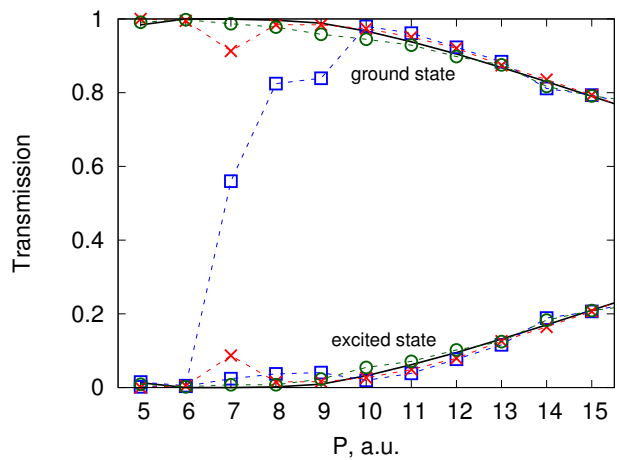}
  \caption{Probability of transmission on the ground and excited
    states with respect to the initial average momentum of the
    distribution located on the ground state: exact quantum (black
    lines), FSSH (red crosses), FSSH+D (blue squares), and PSSH (green
    circles). The dashed lines are not representing results between
    the points but serve as an eye guide.}
  \label{fig:model_2_gs}
\end{figure}

The initial assessment in Ref.~\onlinecite{Shenvi:2009/jcp/124117}
investigated the ability of FSSH+D and PSSH to simulate transmission
of a nuclear wavepacket starting from $\braket{{R}}=-10$. It was shown
that in the case of low initial $\braket{{P}}$, PSSH can model
transmission more accurately that FSSH+D (\fig{fig:model_2_gs}). To
determine whether the failure of FSSH+D simulations resides in the
presence of DBOC, we have redone the nonadiabatic dynamics using FSSH.
As can be seen from \fig{fig:model_2_gs}, FSSH can in fact model
transmission in this model for most cases (FSSH deviates when
$\braket{{P}}\in(6,7)$). In this model DBOC acts as a barrier that
reflects particles with low momenta. DBOC elimination removes the
reflection and allows particles to pass through even at low momenta.
Thus, in this particular model, DBOC should not be simply added to BO
PESs.

The deviation of FSSH for $\braket{{P}}\in(6,7)$ can be explained by
considering that the lowest momentum that permits hopping is ${P}_{\rm
  min}=\sqrt{2M\Delta E_{12}}\approx 6.3$ a.u. Thus, when a classical
particle has ${P}\approx{P}_{\rm min} $, upon hopping to the excited
state it will have a momentum close to zero. Because the BO PESs are
flat, there is no source of acceleration and hopped particles remain
frozen on the excited state.

\begin{figure}[!h]
  \centering
  \includegraphics[width=0.5\textwidth]{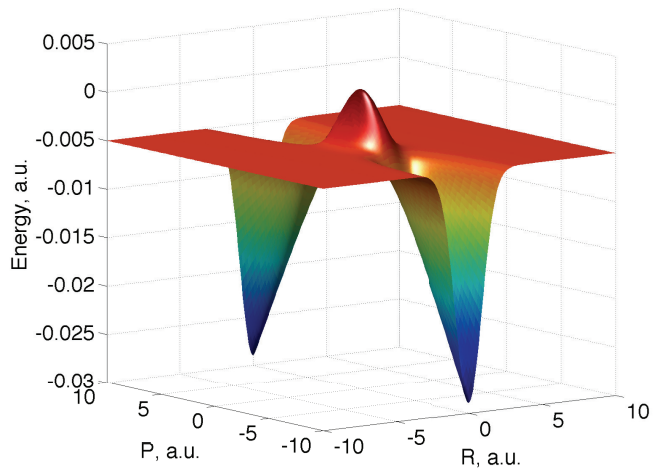}
  \caption{Phase-space ground state potential energy surface without
    the kinetic energy like term, ${E}_{-}^{\rm PS}-{P}^2/(2M)$, for
    the model with flat BO PESs.}
  \label{fig:ps-pes-gs}
\end{figure}

In this model, the difference between curvatures of the potential
energy surfaces in FSSH+D and PSSH methods stems from the square root
term in \eq{eq:PSenergy}. This term partially cancels the repulsive
barrier coming from DBOC for the ground state in PSSH (see
\fig{fig:ps-pes-gs}). Thus the ground state DBOC repulsive barrier is
effectively lower in PSSH than in FSSH+D. This allows PSSH to have
good transmission for PSSH even at low momenta (\fig{fig:model_2_gs}).

\begin{figure}[!h]
  \centering
  \includegraphics[width=0.5\textwidth]{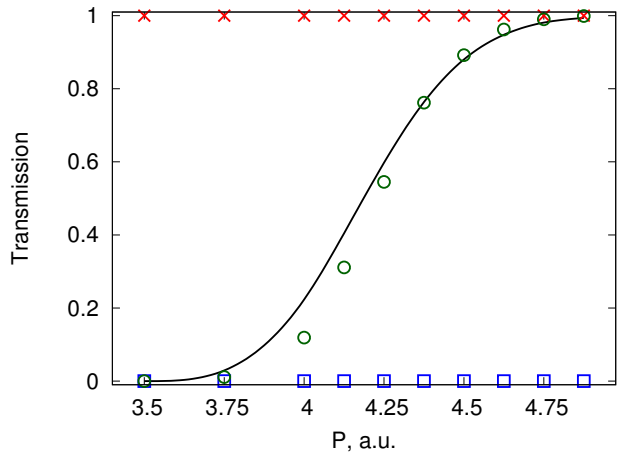}
  \caption{Probability of transmission on the ground state with
    respect to the initial average momentum of the distribution
    located on the ground state: exact quantum (black line), FSSH (red
    crosses), FSSH+D (blue squares), and PSSH (green circles)}
  \label{fig:model_2_gs_low_k}
\end{figure}

Following this logic of compensation there should be a point in the
${P}$-space where the momentum is so low, that DBOC cannot be
compensated by the square root term. To determine how well PSSH
describes the transition from transmitting to reflecting regimes,
simulations for the transmission coefficient have been performed for
low momenta (see \fig{fig:model_2_gs_low_k}). The transmission
coefficients show that for low momenta, DBOC completely repulses
classical particles in FSSH+D, while in FSSH all particles can pass
through the nonadiabatic region. The exact dynamics shows that in the
considered range of momenta, the nuclear wavepacket bifurcates on a
single surface while crossing the nonadiabatic region. PSSH captures
this phenomenon accurately by quantifying adequately the fraction of
the distribution that is transmitted. Note that at this range of
momenta, the nuclear subsystem in all surface hopping variations does
not have enough kinetic energy to hop, therefore, the dynamics is
purely adiabatic.
 
\begin{figure}[!h] 
  \centering
  \includegraphics[width=0.5\textwidth]{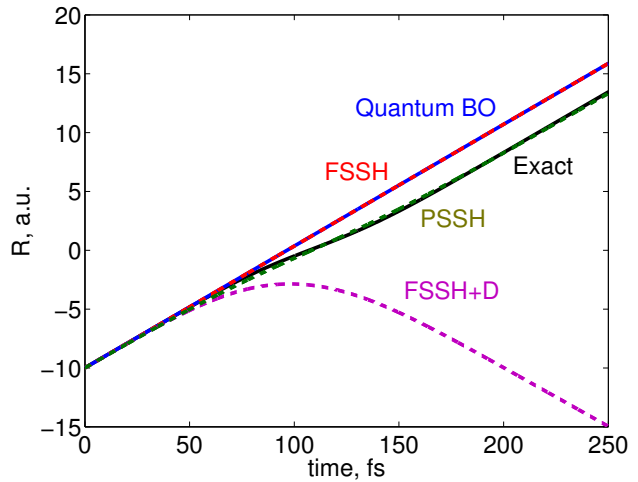}
  \caption{The average position of the nuclear distribution in
    different methods as a function of time: SH variants (dashed
    lines) and quantum calculations (full lines).}
  \label{fig:av_x_per_t}
\end{figure}

It has been noted in Ref.~\onlinecite{Shenvi:2009/jcp/124117} that
momenta of classical particles in PSSH increase while crossing the
nonadiabatic region. However, this does not imply an increase in
velocity. Indeed as shown in \fig{fig:av_x_per_t}, particles in PSSH
actually slow down while crossing the nonadiabatic region on the
ground state. In PSSH, this slow down comes from the square root term
in the potential energy, which has a negative contribution on the
ground state phase-space PES. In the exact quantum dynamics, the
slow-down occurs due to the partial transfer of the wavepacket
population to the excited state. A consequent increase in potential
energy leads to a decrease in the kinetic nuclear energy. In SH
methods, the initial average momentum $\braket{{P}}=5$ a.u. does not
allow hops to occur. In FSSH, the nuclear coordinate does not
experience any force and evolves similarly to the centre of the
nuclear wavepacket on a flat PES in the quantum BO dynamics. Due to
DBOC, FSSH+D overestimates a repulsive character of the electronic
potential. Therefore both FSSH and FSSH+D fail to model accurately the
spatial evolution of the nuclear coordinate when nonadiabatic
couplings are non-negligible. Only PSSH models accurately the
slow-down that a nuclear wavepacket experiences in the true quantum
dynamics in a nonadiabatic region.

\begin{figure}[!h]
  \centering
  \includegraphics[width=0.5\textwidth]{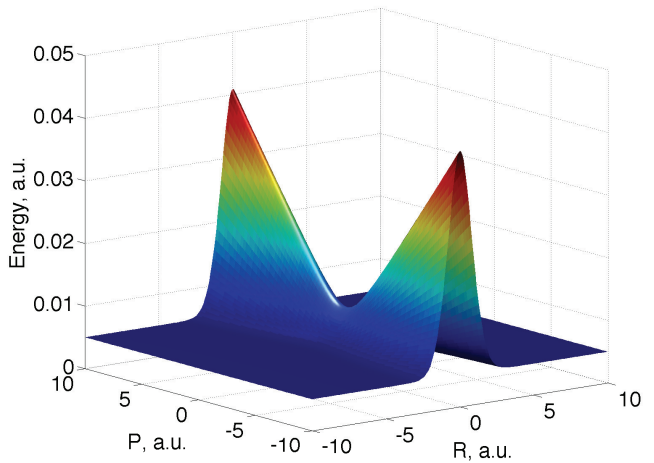}
  \caption{Phase-space excited state potential energy surface without
    the kinetic energy like term, ${E}_{-}^{\rm PS}-{P}^2/(2M)$, for
    the model with flat BO PESs.}
  \label{fig:ps-pes-ex}
\end{figure}

\begin{figure}[!h]
  \centering
  \includegraphics[width=0.5\textwidth]{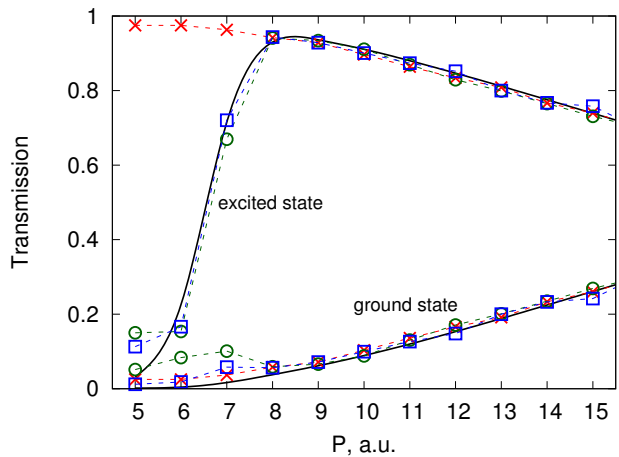}
  \caption{Probability of transmission on the ground and excited
    states with respect to the initial average momentum of the
    distribution located on the excited state: exact quantum (black
    lines), FSSH (red crosses), FSSH+D (blue squares), and PSSH (green
    circles). The dashed lines are not representing results between
    the points but serve as an eye guide.}
  \label{fig:model_2_es}
\end{figure}

In the excited phase-space PES, the square root term adds on to DBOC
and increases the potential energy barrier classical particles need to
surmount in order to pass through nonadiabatic region
(\fig{fig:ps-pes-ex}). By performing the same simulations as
\fig{fig:model_2_gs}, but starting from the excited adiabatic state,
the nuclear wavepacket is repulsed at higher momenta (see
\fig{fig:model_2_es}). Here, FSSH fails completely in the region of
low momenta by showing almost complete transmission. Both FSSH+D and
PSSH reproduce quantum dynamics very well.

\subsection{Avoided crossing}

\begin{figure}[!h]
  \centering
  \includegraphics[width=0.5\textwidth]{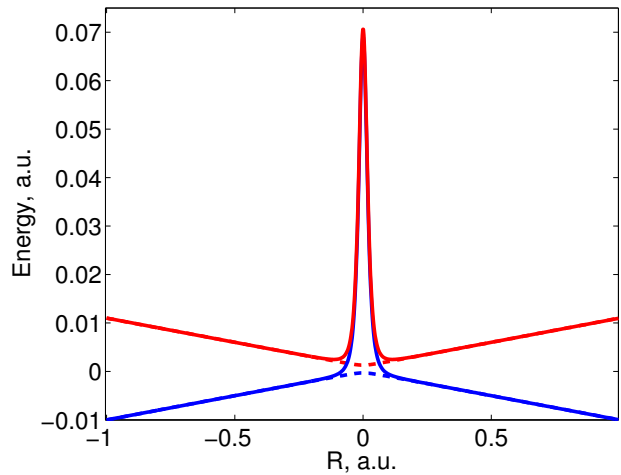}
  \caption{BO (dashed) and adiabatic (solid) PESs of the avoided
    crossing model with $c=3\times10^{-4}$.}
  \label{fig:model_ac_pes}
\end{figure}

While the previous model successfully showed regimes where FSSH,
FSSH+D and PSSH differ, it does not describe situations where BO PESs
come very close to each other. An avoided crossing model allows to
explore such regimes, its diabatic Hamiltonian is \bea H^{\rm D}_{\rm
  AC}=-\frac{1}{2M}\frac{\partial^{2}}{\partial
  {R}^{2}}\mathbf{1}_{2}+\begin{bmatrix}
  -b{R} & c \\
  c & b{R}
\end{bmatrix}, \eea where $b=0.01$ is fixed and $c$ will be varied. In
the adiabatic representation, this model has BO PESs whose lowest
energy gap is $\Delta E_{12}= 2c$ and DBOC is \bea \frac{d_{12}^2}{2M}
= \frac{b^2c^2}{8M(c^2+b^2R^2)^2} \eea (see \fig{fig:model_ac_pes}).
The initial nuclear wavepacket was
$\Psi({R},0)=e^{-4\left({R}-5\right)^2}$. The nuclear mass, $M$, the
number of trajectories, and time-step lengths were taken as in the
model with flat BO PESs.

\begin{figure}[!h]
  \centering
  \includegraphics[width=0.5\textwidth]{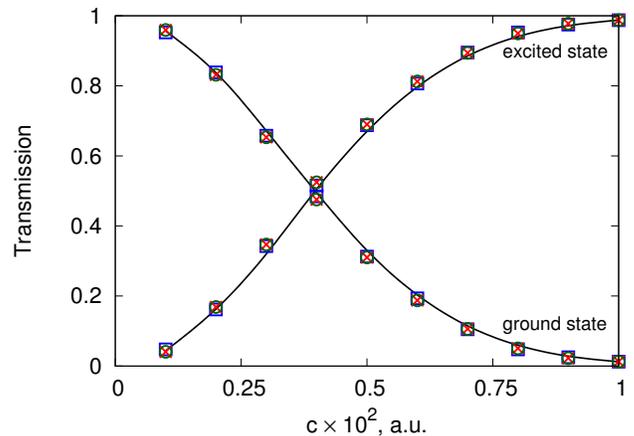}
  \caption{Probability of transmission on the ground state with
    respect to the diabatic coupling constant, $c$: exact quantum
    (black line), FSSH (red crosses), FSSH+D (blue squares), and PSSH
    (green circles). The initial nuclear distribution is on the
    excited state.}
  \label{fig:avoid_cross_es}
\end{figure}

\begin{figure}[!h]
  \centering
  \includegraphics[width=0.5\textwidth]{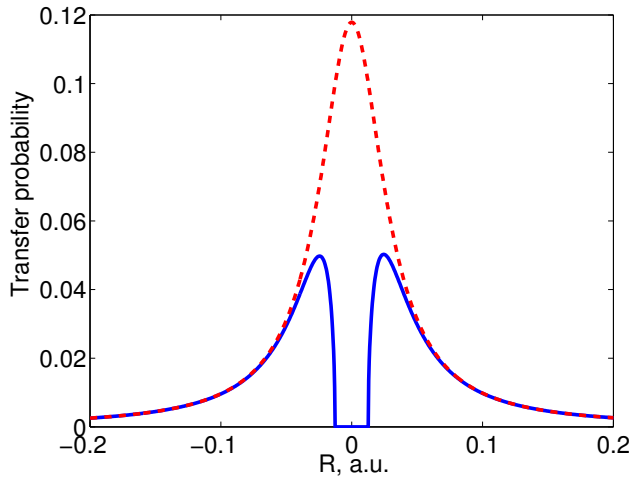}
  \caption{Transfer probability ($d_{12}P/M$) as a function of
    position for the avoided crossing model with $c=3\times10^{-4}$:
    FSSH (red dashed) and FSSH+D (solid blue). $P$ as a function of
    $R$ is evaluated using $\sqrt{2M\max(E_0-E_{+}(R),0)}$, where $E_0
    = -bR_0$ is the initial energy with $R_0=-5$ a.u. and $E_{+}(R)$
    are excited state PESs with and without DBOC.}
  \label{fig:fssh_d_actp}
\end{figure}

The property of interest is the probability for the nuclear wavepacket
starting from the excited state to transfer to the ground state. When
the diabatic coupling constant, $c$, is high, the wavepacket is
expected to remain on the upper adiabatic state for the entire
simulation while for low diabatic constant, a nearly complete transfer
to the lower adiabatic state is envisioned. Figure
\ref{fig:avoid_cross_es} presents results for a range of $c$'s that
corresponds to a range of DBOC maximum energies of $6.3\times10^{-3}$
-- $6.3\times 10^{-5}$ a.u. These energies are much smaller than the
kinetic energy that the wave-packet gains at $R=0$, $5\times10^{-2}$
a.u., and thus, all SH variations model nuclear dynamics accurately
(\fig{fig:avoid_cross_es}).
\begin{figure}[!h]
  \centering
  \includegraphics[width=0.5\textwidth]{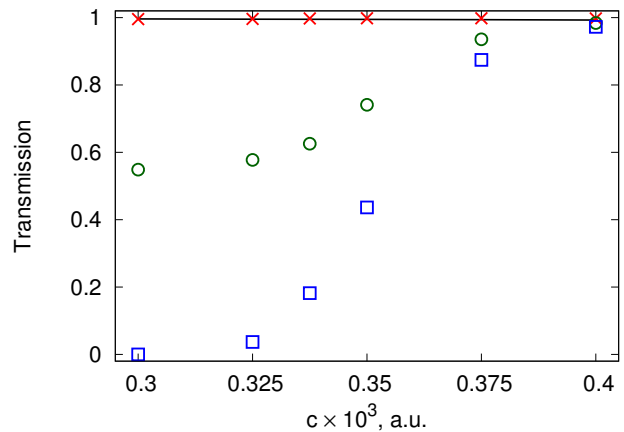}
  \caption{Probability of transmission on the ground and excited
    states with respect to diabatic coupling constant, $c$: exact
    quantum (black lines), FSSH (red crosses), FSSH+D (blue squares),
    and PSSH (green circles). The initial nuclear distribution is on
    the excited state.}
  \label{fig:avoid_cross_es_critical}
\end{figure}
Decreasing $c$ to values where DBOC maxima are comparable or higher
than the kinetic energy of the nuclear wave-packet at $R=0$ separates
all SH methods, see \fig{fig:avoid_cross_es_critical}. In both FSSH+D
and PSSH, the system transfer to the ground state is significantly
inhibited by increasing DBOC. Qualitatively, for both methods the
reason for this deviation is similar but it is easier to illustrate it
in the FSSH+D case (see \fig{fig:fssh_d_actp}). In FSSH+D, a repulsive
DBOC potential reduces nuclear momentum of a particle and as a
consequence the nonadiabatic transfer probability to zero before the
particle reaches the intersection $R=0$. This allows the nonadiabatic
transfer to take place only before the intersection where the particle
will not have enough kinetic energy to overcome the DBOC induced
barrier on the ground state.

PSSH suffers less from the DBOC inclusion
(\fig{fig:avoid_cross_es_critical}) because of two reasons: First, if
the non-adiabatic transfer happens to the ground state, DBOC can be
compensated there by the $d_{12}P/M$ term. Second, PSSH particles have
generally larger velocities on the excited state in the nonadiabatic
region. In this region the difference between BO PES becomes
negligible ($E_{2}(R)-E_{1}(R)\approx0$) and the nuclear velocity in
PSSH can be approximated as \bea
\label{eq:velocity}
\dot{R}_{\pm}=
\frac{P}{M}\left(1\pm\left|\frac{d_{12}}{P}\right|\right). \eea Thus,
nuclear DOF can experience an acceleration if they are on the
phase-space excited state and a deceleration if they are on the
phase-space ground state. For classical particles the DBOC effect will
reduce $P$. However, due to $d_{12}$ the $P$ reduction does not lead
to velocity reduction. In other words, in PSSH, nuclei can use
$d_{12}$ to overcome a part of the DBOC repulsion.

On the other hand FSSH correlates with the exact dynamics and shows
complete transfer. In the absence of DBOC, nothing prevents classical
particles in FSSH from accessing the region of strong nonadiabatic
coupling and hopping to the ground state (\fig{fig:fssh_d_actp}). The
final outcome will not depend on whether a hop taken place before or
after the intersection because the ground state does not have a DBOC
induced barrier. Thus for weakly diabatically coupled avoided crossing
models, FSSH surpasses both PSSH and FSSH+D in describing excited
state dynamics.

\begin{figure}[!h]
  \centering
  \includegraphics[width=0.5\textwidth]{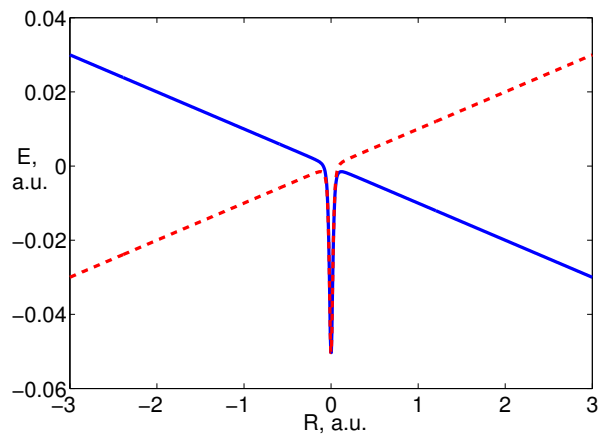}
  \caption{Diabatic surfaces (solid blue and dashed red) for the
    avoided crossing model when DBOC is subtracted from the
    Hamiltonian in the adiabatic representation,
    $c=3.5\times10^{-4}$.}
  \label{fig:m_dboc}
\end{figure}

\begin{figure}[!h]
  \centering
  \includegraphics[width=0.5\textwidth]{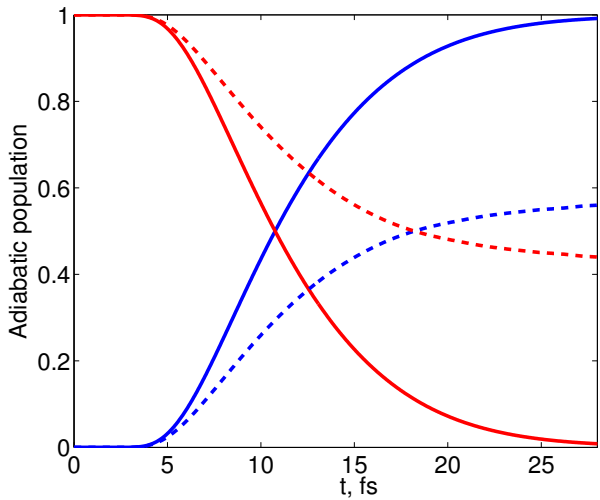}
  \caption{Dynamics of the adiabatic population starting from the
    initial wavepacket on the excited state, $M=200$ a.u.,
    $c=3.5\times10^{-4}$: red and blue are populations of the excited
    and ground states, solid and dashed are with and without DBOC,
    respectively.}
  \label{fig:m_dboc_dyn}
\end{figure}

Interestingly, in the small $c$ case, interpreting DBOC as a repulsive
potential in quantum dynamics is incorrect because in the nonadiabatic
region ($R\approx 0$) NAC becomes very large \bea d_{12} =
\frac{bc}{2(c^2+b^2R^2)}\rightarrow \frac{b}{2c},\quad R\rightarrow 0
\eea and thus the adiabatic surface interpretation of the dynamics is
misleading. Instead, the simplest quantum dynamical picture emerges in
the diabatic representation where for very small $c$'s dynamics is
almost fully confined to a single diabatic surface. In the diabatic
representation, DBOC does not appear and the absence of any other
repulsive potentials on the diabats illustrates that there is a
complete cancellation of diagonal and off-diagonal derivative coupling
terms when one goes from the adiabatic representation to the diabatic
one. Moreover, if one subtracts DBOC from the adiabatic nuclear
Hamiltonian [\eq{eq:molHmodPES}] the transfer dynamics becomes slower
and less efficient. Thus, effectively, removing DBOC introduces the
repulsive potential in the quantum dynamics. To understand this, it is
instructive to transform the adiabatic Hamiltonian without DBOC to the
diabatic representation\footnote{The adiabatic-to-diabatic
  transformation is exactly the same as without subtracting since DBOC
  is the diagonal term that has the same value for both states.} where
subtracting the DBOC leads to two dips on the diabats at the point of
their intersections (\fig{fig:m_dboc}). These dips give rise to the
over-barrier reflection of the wave-packet traveling on a diabat and
thus reduces the efficiency of passing the crossing point
(\fig{fig:m_dboc_dyn}). This is purely quantum effect and it will be
lost when classical mechanics is used on the same potential.

\subsection{Conical intersections}

Conical intersections are ubiquitous in molecular systems and allow
for ultra-fast transfer between electronic states. At the exact point
of intersection, electronic states are degenerate and give rise to an
infinitely large DBOC, which have been shown to decrease the rate of
electronic transitions in FSSH+D.\cite{Gherib:2015/jctc/11} Analysis
of the interplay between DBOC and other nonadiabatic terms in fully
quantum dynamics for CIs is complicated by appearance of a nontrivial
geometric phase and is provided in
Ref.~\onlinecite{Ryabinkin:2014/jcp/140}. It was found that for CIs,
DBOC is only compensated by other terms when the geometric phase is
included. Without geometric phase, DBOC creates repulsive potential
for the quantum nuclear wavepacket, therefore, since SH methods do not
have geometric phase for the nuclear wave-function, they also
experience DBOC as a repulsive potential even in a greater extent
because classical particles cannot tunnel under DBOC.

To determine whether PSSH can model population dynamics through CIs,
we consider the 2D-LVC model \bea
\label{eq:dia_LVC_Hamil}
H^{\rm D}_{\rm LVC}=T_{2D}\mathbf{1}_{2}+\begin{bmatrix}
  V_{11} & V_{12} \\
  V_{12} & V_{22}
\end{bmatrix}, \eea where
\begin{align}
  \label{eq:diabatic2dlvc_1}
  V_{11}&=\frac{1}{2}\left[\omega_{1}^2\left(x+\frac{a}{2}\right)^2+\omega_{2}^2y^2+{\Delta}\right],  \\
  \label{eq:diabatic2dlvc_2}
  V_{22}&=\frac{1}{2}\left[\omega_{1}^2\left(x-\frac{a}{2}\right)^2+\omega_{2}^2y^2-{\Delta}\right], \\
  \label{eq:diabatic2dlvc_3}
  V_{12}&=cy .
\end{align}
This diabatic model corresponds to two paraboloids shifted in space in
the $x$-direction by $a$ and in energy by $\Delta$. Three molecular
systems whose ultrafast excited state dynamics is well represented
with 2D-LVC have been investigated: bis(methylene) adamantyl cation
(BMA),\cite{Blancafort:2005/jacs/3391} butatriene
cation\cite{Koppel:1984/acp/59,Cederbaum:1977/cp/169,Cattarius:2001/jcp/2088,Sardar:2008/pccp/6388,Burghardt:2006/mp/1081,Gindensperger:2006/jcp/144103}
and
pyrazine.\cite{Seidner:1993/cpl/117,Woywod:1994/jcp/1400,Sukharev:2005/pra/012509}
Their 2D-LVC parameters are given in Table \ref{tab:BMA-param}.

\begin{table}[!h]
  \caption{Parameters of the 2D-LVC
    Hamiltonian, \eq{eq:dia_LVC_Hamil} for the three CI systems.} 
  \label{tab:BMA-param}
  \centering
  \begin{ruledtabular}
    \begin{tabular}{@{}lccccr@{}}
      \multicolumn{1}{c}{$\omega_1$} & $\omega_2$ & $a$ & $c$ &
      \multicolumn{1}{c}{$\Delta$}  \\ \hline
      \multicolumn{6}{c}{ Bis(methylene) adamantyl cation} \\
      $7.743\times10^{-3}$ & $6.680\times10^{-3}$ & 31.05 &
      $8.092\times 10^{-5}$ & 0.000 & \\[1ex]
      \multicolumn{6}{c}{ Butatriene cation} \\
      $9.557\times10^{-3}$ & $3.3515\times10^{-3}$  & 20.07   &
      $6.127\times 10^{-4}$ & 0.020 \\[1ex]
      \multicolumn{6}{c}{ Pyrazine} \\
      $3.650\times10^{-3}$ & $4.186\times10^{-3}$ & 48.45 & $4.946\times
      10^{-4}$ & 0.028 
    \end{tabular}
  \end{ruledtabular}
\end{table}

MQC simulations are done using 2000 trajectories and 0.05, 0.01, and
0.001 fs time-steps for FSSH, FSSH+D, and PSSH, respectively.
Similarly to FSSH, in PSSH each trajectory carries both an electronic
wavefunction and an active electronic surface. However, in PSSH these
quantities correspond to the phase-space basis and PESs, which are
identical to their adiabatic counterparts far from the nonadiabatic
region, but differ from them when adiabatic states become coupled. To
model adiabatic population dynamics using PSSH, one is faced with the
following problem: How to use phase-space information to calculate the
adiabatic populations ?

A straightforward procedure consists in rotating the electronic
wavefunction to the adiabatic representation using a unitary matrix
and taking absolute squares of the complex amplitudes to obtain the
adiabatic populations. An alternative method consists in ignoring the
electronic wavefunction and decomposing the active phase-space surface
into adiabatic state weights. Both methods were used in the following
simulations, and we will denote population calculations based on the
amplitudes of electronic wavefunctions as PSSH-A and based on the
active phase-space surface as PSSH-S.

\begin{figure}[!h]
  \centering
  \includegraphics[width=0.5\textwidth]{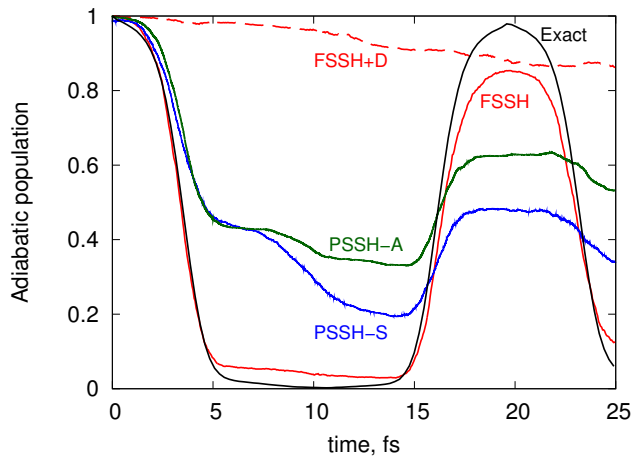}
  \caption{Excited state adiabatic population dynamics for BMA cation
    in different methods.}
  \label{fig:bma_es}
\end{figure}

\begin{figure}[!h]
  \centering
  \includegraphics[width=0.5\textwidth]{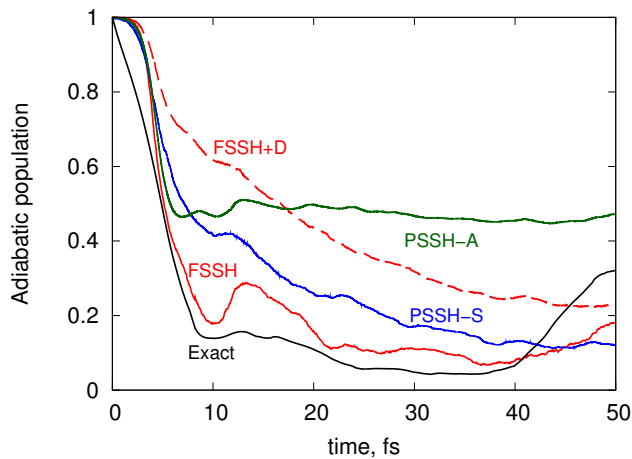}
  \caption{Excited state adiabatic population dynamics for butatriene
    cation in different methods.}
  \label{fig:buta_es}
\end{figure}

\begin{figure}[!h]
  \centering
  \includegraphics[width=0.5\textwidth]{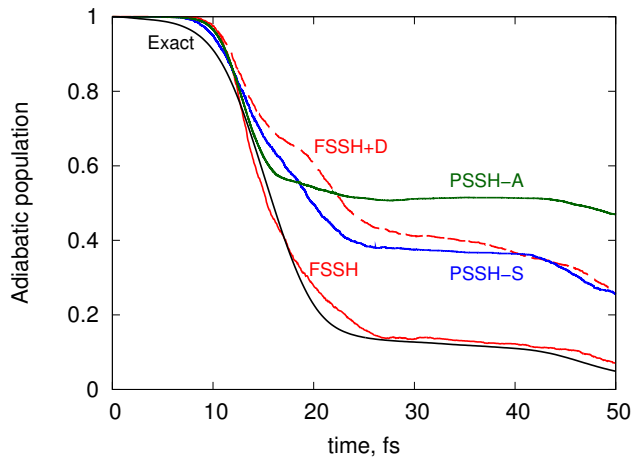}
  \caption{Excited state adiabatic population dynamics for pyrazine in
    different methods.}
  \label{fig:pyra_es}
\end{figure}

For all CI models, PSSH nonadiabatic dynamics is in a worse agreement
with the exact one than that of FSSH (see
Figs.~\ref{fig:bma_es}-\ref{fig:pyra_es}). However, PSSH clearly
outperforms FSSH+D. In the cases of BMA (\fig{fig:bma_es}) and
pyrazine (\fig{fig:pyra_es}), there is only partial population
transfer within the considered time span. This failure of PSSH does
not stem from the procedure used to convert phase-space electronic
information into adiabatic populations, but rather from the presence
of DBOC in phase-space PESs. DBOC repulses classical particles away
from regions where hops are probable.

In light of the results of \fig{fig:avoid_cross_es_critical}, the
failure of PSSH is justified. In system with CIs, most classical
particles never go through the CI, instead, the majority evolve on
PESs that resemble avoided crossings. Particles travelling in regions
of low diabatic couplings will experience greater DBOC and will be
pushed away from nonadiabatic regions.

\section{Conclusions}

We systematically assessed the inclusion of DBOC in surface hopping
methods for various one- and two-dimensional nonadiabatic models. It
was found that for DBOC to affect dynamics its energy scale must be
larger or comparable with that of the nuclear kinetic energy. In cases
when DBOC is large the off-diagonal NACs are also significant. This
relation makes improving BO PESs by adding DBOC to them less appealing
because a PES picture is only adequate when corresponding couplings
are small. Inherently, all surface hopping methods use classical
mechanics to describe the nuclear motion within a PES and stochastic
treatment of inter-surface couplings. Therefore the best
representation for these methods needs to have low overall couplings
between PESs.

When DBOC is simply added to BO PESs, the FSSH+D approach, it always
brings a repulsive potential that slows down classical particles and
thus makes nonadiabatic transitions less probable. For the dynamics on
the excited state of the one-dimensional model with flat BO PESs this
behaviour is in accord with the exact quantum nuclear dynamics.
However, for the ground state dynamics of the same system and excited
state dynamics of the avoided crossing and conical intersection models
FSSH+D overestimates the effect of the DBOC repulsion and deviates
qualitatively from the exact dynamics.

More advanced treatment of DBOC via the PSSH approach operates with
phase-space PESs that account for some interplay between DBOC and
off-diagonal NACs. For all cases, PSSH performed better than FSSH+D,
this can be related to the DBOC compensation by NACs for the ground
state dynamics and NACs contribution to velocity enhancements in
nonadiabatic regions that allows particle to advance further on the
excited state in spite of the DBOC repulsion. However, in very weakly
coupled avoided crossing and conical intersection problems, PSSH
performance was worse than the that of the original FSSH method
without DBOC. We attribute this to difficulty of capturing the correct
interplay between DBOC and NACs at a very localized nonadiabatic
regions appearing in these problems. The full quantum formalism is
capable of treating this interplay mainly because it involves quantum
nuclear wavefunctions. Also, modelling a very weakly diabatically
coupled avoided crossing system revealed that a repulsive potential
consideration of DBOC that appears in surface hopping treatment is
qualitatively incorrect. In this case DBOC is responsible for
providing smooth diabatic surfaces, and its removal leads to diabatic
surfaces with a prominent over-barrier reflection for a quantum
nuclear wavepacket.

Applying PSSH on large molecular systems in conjunction with
electronic structure methods poses a future challenge. The algorithm
requires not only computing first and second-order nonadiabatic
couplings but also their gradients with respect to nuclear
coordinates. The latter are necessary to compute the forces acting on
nuclei evolving on phase-space PESs. Furthermore, for weakly
diabatically coupled avoided crossing and conical intersection
problems, there is no advantage in using PSSH instead of FSSH. The
former class of systems frequently appear in simulations of long range
charge and energy transfers and have been referred in literature as
trivial unavoided crossings.\cite{Meek:2014jc,Wang:2014jf} Thus,
considering relatively small range of systems where adding DBOC could
improve current surface hopping approaches and additional
computational expenses for DBOC evaluation it is not advisable to
incorporate this quantity in the mixed quantum-classical calculations.

Current findings are also in accord with our recent work on the
quantum classical Liouville equation
(QCLE)\cite{Ryabinkin:JCP/2014/084104} which is a more advanced and
rigorously derivable mixed quantum-classical
approach.\cite{Kapral:2006kw} Two main steps in the QCLE derivation
are a projection to adiabatic electronic states and a Wigner
transformation of nuclear coordinates. These two steps do not commute
and their different orders give rise to two different methods:
Adiabatic-then-Wigner
(AW)\cite{Ando:2002/cpl/240,Horenko:2002/jcp/11075} and
Wigner-then-Adiabatic (WA)\cite{Kelly:2010/jcp/084502} QCLEs. Although
two methods perform in many instances
similarly,\cite{Ando:2003/jcp/10399} based on analysis of conical
intersection models and associated geometric phase effects it was
found that only WA-QCLE is mathematically well defined
approach.\cite{Ryabinkin:JCP/2014/084104} Interestingly, DBOC does not
appear in WA-QCLE, but it is a part of AW-QCLE. Recently, FSSH
approach has been connected to WA-QCLE
method\cite{Subotnik:2013jw,Kapral:CP/2016} and in light of this
connection it is natural that FSSH should not include DBOC.

\section{Acknowledgements}

A.F.I. would like to thank Neil Shenvi for helpful discussions and
acknowledges funding from the Natural Sciences and Engineering
Research Council of Canada (NSERC) through the Discovery Grants
Program and the Alfred P. Sloan Foundation. R.G. would like to
acknowledge funding from the Queen Elizabeth II Graduate Scholarship
in Science and Technology.


%

\end{document}